\documentclass[pre,twocolumn]{revtex4-1}

\usepackage{amsmath} 
\usepackage{graphicx}
\begin{document}

\title{Efficiency estimation for an equilibrium version of Maxwell refrigerator}
\author{Toby Joseph}
\email{Electronic mail: toby@goa.bits-pilani.ac.in}
\author{V. Kiran}
\affiliation{Department of Physics, BITS Pilani K K Birla Goa Campus, Zuarinagar 403726, Goa, India}
\date{6th June, 2020}

\begin{abstract}
Maxwell refrigerator as a device that can transfer heat from a cold to hot temperature reservoir making 
use of information reservoir was introduced by Mandal et al. \cite{Mandal2013a}. The model has a two 
state demon and a bit stream interacting with two thermal reservoirs simultaneously. We work out a 
simpler version of the refrigerator where the demon and bit system interact with the reservoirs separately 
and for a duration long enough to establish equilibrium. The efficiency, $\eta$, of the device when working 
as an engine as well as the coefficient of performance (COP) when working as a refrigerator are calculated. 
It is shown that the maximum efficiency matches that of a Carnot engine/refrigerator working between the 
same temperatures, as expected. The COP at maximum power decreases as $\frac{1}{T_h}$ when 
$T_h >T_c \gg \Delta E$ ($k_B = 1$), where $T_h$  and $T_c$ are the temperatures of the hot and cold 
reservoirs respectively  and $\Delta E$ is the level spacing of the demon. $\eta$ at maximum power of the 
device, when working as a heat engine, is found to be $\frac{T_h}{0.779 + T_h}$ when $T_c \ll \Delta E$ 
and $T_h \gg \Delta E$.
\end{abstract}

\maketitle

Maxwell \cite{ClerkMaxwell:1871:TH} introduced an `intelligent' demon that could discern the speed of 
molecules in two chambers connected by a trap door. Using this information and operating a trapdoor
connecting the two chambers, the demon can transfer heat from cold to hot body, thus apparently violating 
the second law of thermodynamics. The discussions on this thought experiment raged through most of the 
twentieth century \cite{Smoluchowski1912,Szilard:1929:EET,Feynman:1963:FLP,Brillouin1951,OPenrose1979} 
and found a full resolution with the works of Landauer and Bennett \cite{Landauer1961,Bennett1982,
Bennett1985}. Crucial to the understanding of the problem is the fact that the demon needs to 
{\it record} the outcome of the measurement experiment and this involves increasing the information 
entropy of the memory registers. Resetting these memory elements leads to generation of entropy, which 
if accounted for explains the validity of second law \cite{Maruyama2009}.

Recently there has been a renewed interest in physics related to Maxwell demon to better understand 
the inter-relation between information and thermodynamics in varied systems and devices. These include 
experimental work where Maxwell Demon like set up has been implemented and studied \cite{Berut2012,
Toyabe2010, Masuyama2018, Paneru2020}, exploring the relevance of these ideas in biomolecular systems 
\cite{Ito2015,Tu11737} and molecular machines \cite{Euan2015}. Theoretical studies on models 
of Maxwell's demon like systems, both autonomous and with feedback loops 
\cite{Quan2006,Abreu2011,Mandal2012,Mandal2013a,Barato2014a,Ribezzi-Crivellari2019}, has been of
carried out recently. The analysis of these models in the light of generalized fluctuation theorems in 
non-equilibrium thermodynamics has been of recent focus \cite{Seifert2012,Kim2007,Sagawa2008,Annby2020}.

One of the earliest exactly solvable model of Maxwell refrigerator (MR) was proposed by Mandal et al. 
\cite{Mandal2013a}. It is an autonomous device consisting of a two level system (the demon) and a 
bit stream, that serves as a memory element, interacting simultaneously with two thermal reservoirs (Refer
Fig. (1) in \cite{Mandal2013a}). By solving for steady state of the demon using the master equation for the 
demon and bit system, it was shown that in certain regime of initial probability distribution of bit states the 
device acts as a refrigerator. In this work we present a related but simpler model, where the demon 
(two level system) first interacts with the hot reservoir and subsequently the demon and current bit in the 
bit stream interacts with the cold reservoir. After proving the validity of the second law for this system 
and working out various modes of operation of the device, we estimate the efficiency and COP of the device 
in these modes. Specifically we study the efficiency/COP in the reversible case and at maximum power.
 
The model consists of, as in the case of MR, a moving bit stream, a two level demon and two thermal reservoirs at
temperatures $T_h$ and $T_c$. The primary difference between the new model and the MR model is in that 
we assume the bit plus demon system interacts with the two thermal reservoirs separately  
(see Fig. (1) given in the supplementary). Further, the interaction duration is taken to be large enough such 
that the system comes to equilibrium with the two reservoirs alternatively. We shall assume that the incoming bits 
are uncorrelated and the probability of a bit being in zero state is $p_0$. The rules for the transition when in 
contact with the two thermal reservoirs are such that while in contact with the hot reservoir, the bit states are 
left unaltered and while in contact with the cold reservoir transitions are allowed only between states $0d$ to 
$1u$. The ratio of rates given by $\frac{p_{1u \rightarrow 0d}}{p_{1u \leftarrow 0d}} =  
e^{\beta_c \Delta E}$, where $\Delta E \equiv E_u - E_d$ being the energy difference between the two 
states of the demon and $\beta_c = \frac{1}{T_c}$ is the inverse temperature (we are using the convention that
the Boltzmann constant, $k_B = 1$).

We will now solve for the probability of the outgoing bit to be $1$. Consider the demon and the 
current bit in contact with the cold reservoir. Since the demon state would be determined by its contact with the 
hot reservoir with which it was in equilibrium, the probability that the state of the demon is down will be
 $\frac{1 + \sigma}{2}$ and for it do be in the up state will be $\frac{1 - \sigma}{2}$, where 
$\sigma = \tanh(\frac{\beta_h \Delta E}{2})$ with $\beta_h = \frac{1}{T_h}$ . After the interaction of the 
demon and bit combined system with the cold reservoir there are three possible ways for the current bit to end 
up in state $1$: (i) The current bit at the beginning was $1$ and the demon state at the beginning was down. 
The probability for this event is $p_1 \frac{(1 + \sigma)}{2}$, where $p_1 = 1 - p_0$ is the probability for the 
incoming bit to be one. (ii) The current bit at the beginning was $0$ and the demon state at the beginning 
was down. And the end state after interaction with the cold reservoir is $1u$. The probability for this event is 
$p_0 \frac{(1 + \sigma)}{2} \frac{(1 - \omega)}{2}$, where $\omega = \tanh(\frac{\beta_c \Delta E}{2})$ 
and $\beta_c = \frac{1}{T_c}$. (iii) The current bit at the beginning was $1$ and the demon state at the 
beginning was up. And the end state after interaction with the cold reservoir is $1u$. The probability for this 
event is $p_1 \frac{(1 - \sigma)}{2} \frac{(1 - \omega)}{2}$. Adding all the probabilities above we find 
the probability, $p_1'$, for the outgoing bit to be $1$ is
\begin{equation}
p_1' = p_1 \frac{(1 + \sigma)}{2} + p_0 \frac{(1 + \sigma)}{2} \frac{(1 - \omega)}{2} 
+ p_1 \frac{(1 - \sigma)}{2} \frac{(1 - \omega)}{2}
\label{p1p}
\end{equation}

Since any absorption of heat from the cold reservoir comes with a flip of the current bit from $0$ state to 
$1$ state, the average heat transferred per cycle from the cold reservoir to the hot reservoir will be,  
$Q = \Phi \;\Delta E$, where $\Phi \equiv p_1' - p_1$ is the increase in the occurrence of $1$s in the outgoing 
bit stream compared to the incoming stream. In the current model, the outgoing bits are not
correlated as the demon equilibrate with the hot reservoir before it comes in contact with the cold reservoir
again. This ensures that any memory of its state after the interaction with the cold reservoir  (along with the
bit state) in the previous cycle is lost. Putting in the value of $p_1'$ from Eq. (\ref{p1p}) 
(defining $\delta \equiv p_0 - p_1$, $\epsilon \equiv \frac{\omega - \sigma}
{1 - \omega \sigma}$), and simplifying one gets
\begin{equation}
\Phi = \frac{(\delta - \epsilon)(1 - \sigma \omega)}{4} \;.
\label{phi}
\end{equation}
As in the original $MR$ model \cite{Mandal2013a}, one sees that there is a transfer of heat from cold to hot 
reservoir provided $\delta > \epsilon$. Parameter $\delta$ measures the capacity of the information reservoir 
(the incoming bit stream) to transfer heat from cold to hot reservoir and the parameter $\epsilon$ is a 
measure of the temperature difference between the reservoirs. It is to be noted that the expression for 
$\Phi$ in the current model is different from the one obtained for the  MR model \cite{Mandal2013a}, 
both for the general case as well as for the special case where the interaction rate with the hot 
reservoir goes to infinity. 

\begin{figure}
\begin{center}
\includegraphics[height = 6cm]{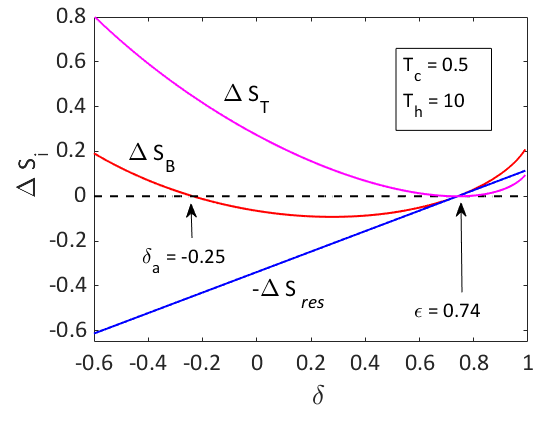}
\caption{Various entropy terms as a function of $\delta$ for $T_c = 0.5$ and $T_h = 10$. This plot 
gives the generic behavior of these terms as function of $\delta$. Above $\delta > \epsilon$, 
$-\Delta S_{res}$ is positive indicating that heat is transferred from the cold reservoir to the hot reservoir. 
For $\delta_a < \delta < \epsilon$, heat is transferred from hot to cold reservoir, but the entropy change of 
bits is negative. This implies that in this regime the demon and bits system is working as an eraser.}
\label{fig.1}
\end{center}
\end{figure}
The seeming violation of the second law, in that one is able to transfer heat from cold body
to hot body, is corrected for provided one considers the increase in information entropy of the bit stream. 
The change in information content of the bits is given by
\begin{equation}
\Delta S_B =  p_1 \ln(p_1) + p_0 \ln(p_0) - p_1' \ln(p_1') - p_0' \ln(p_0')
\label{sbit}
\end{equation}
where the primed quantities refer to the probabilities associated with the outgoing bit stream.
By Landauer's principle this is the minimum entropy that will be generated as one resets the outgoing 
memory bits to the same probability distribution as the incoming ones. The entropy change in the thermal 
reservoirs is
\begin{equation}
\Delta S_{res} = \Phi \; (\beta_h - \beta_c)
\label{sres}
\end{equation}
where we have put $\Delta E = 1$, defining the energy scales with respect to the level spacing.
The sum of the above two entropies (given in Eq. (\ref{sbit}) and Eq. (\ref{sres}) )is the average change in entropy 
of the system per cycle and can be written as
\begin{eqnarray}
\Delta S_{T} &=& \frac{1 - \delta}{2} {\ln}\frac{1 - \delta}{1 - \delta + 2 \Phi} 
+ \frac{1 + \delta}{2} {\ln}\frac{1 + \delta}{1 + \delta - 2 \Phi} \nonumber \\
&+& \Phi \;{\ln}\frac{(1 + \delta - 2\Phi)(1 - \epsilon)}{(1 - \delta + 2 \Phi)(1 + \epsilon)}
\label{stot}
\end{eqnarray}
where we have used $p_0' = 1 - p_1'$,  $p_1' = \Phi + p_1$, $p_1 = 1 - p_0$ and $p_0 = \frac{\delta + 1}{2}$.

It can be shown that $\Delta S_T$ is larger than or equal to zero for all parameter values, 
thus validating the second law. The first two terms in Eq. (\ref{stot}) is the Kullback-Leibler 
divergence between probability distribution of the incoming and the outgoing bits, $D_{KL}(p||p')$.
The positivity of this quantity (by Gibb's identity) ensures that the sum of first two term is always
larger than or equal to zero. The third term ($T3$) in the RHS of the equation above can be shown
to be always larger than or equal to zero by looking at three separate cases: $\delta = \epsilon$, 
$ \epsilon < \delta \le 1$ and $ -1 \le \delta < \epsilon$. When $\delta = \epsilon$, $T3$ is zero since $\Phi = 0$. 
When $\epsilon < \delta \le 1$, $\Phi > 0$ and it can be shown that the argument inside the logarithm is larger 
than $1$ (see supplementary). And when $-1 \le \delta < \epsilon$, $\Phi < 0$ and it can be shown that the 
argument inside the logarithm is between $0$ and $1$. These observations ensure that $T3$ is 
always larger than or equal to zero thus proving that $\Delta S_{T} \ge 0$. Note that the process is 
reversible for the case when $\delta = \epsilon$ and there is no heat exchange between the thermal
reservoirs for this case.

In Fig.~\ref{fig.1}, the variation of $\Delta S_B$ and $-\Delta S_{res}$ are plotted as a function 
of $\delta$ for $T_c = 0.5$ (in units of $\frac{\Delta E}{k_B}$) and $T_h = 10$. Note that $\Delta S_B$ 
is zero at $\delta = \epsilon$ and it has another zero for a lower value of $\delta$, which we refer to as 
$\delta_a(\omega, \epsilon)$. Since $\Delta S_T = \Delta S_B +\Delta S_{res} \ge 0$ and $\Delta S_T = 0$ 
at $\delta = \epsilon$, the straight line curve corresponding to $-\Delta S_{res}(\delta)$ is tangential to the 
curve $\Delta S_B (\delta)$ at the point $\delta = \epsilon$. The regime $\epsilon < \delta \le 1$ is where 
the device works as a refrigerator, since in this region $Q$ is positive. In the regime 
$\delta_a < \delta < \epsilon$, $\Delta S_B < 0$ and hence there is an erasure of the incoming bits on 
the average. Assuming a Szilard type reversible mechanism to convert the information entropy gained 
into work (and thus setting the probability distribution of bits to its incoming value), one can consider the 
$\Delta S_B < 0$ region as one where the system is working as a heat engine. When $-1 \le \delta < \delta_a$, 
$\Delta S_B > 0$ and $Q < 0$ implying that the system is neither acting as a refrigerator nor a heat engine.

\begin{figure}
\begin{center}
\includegraphics[height = 6cm]{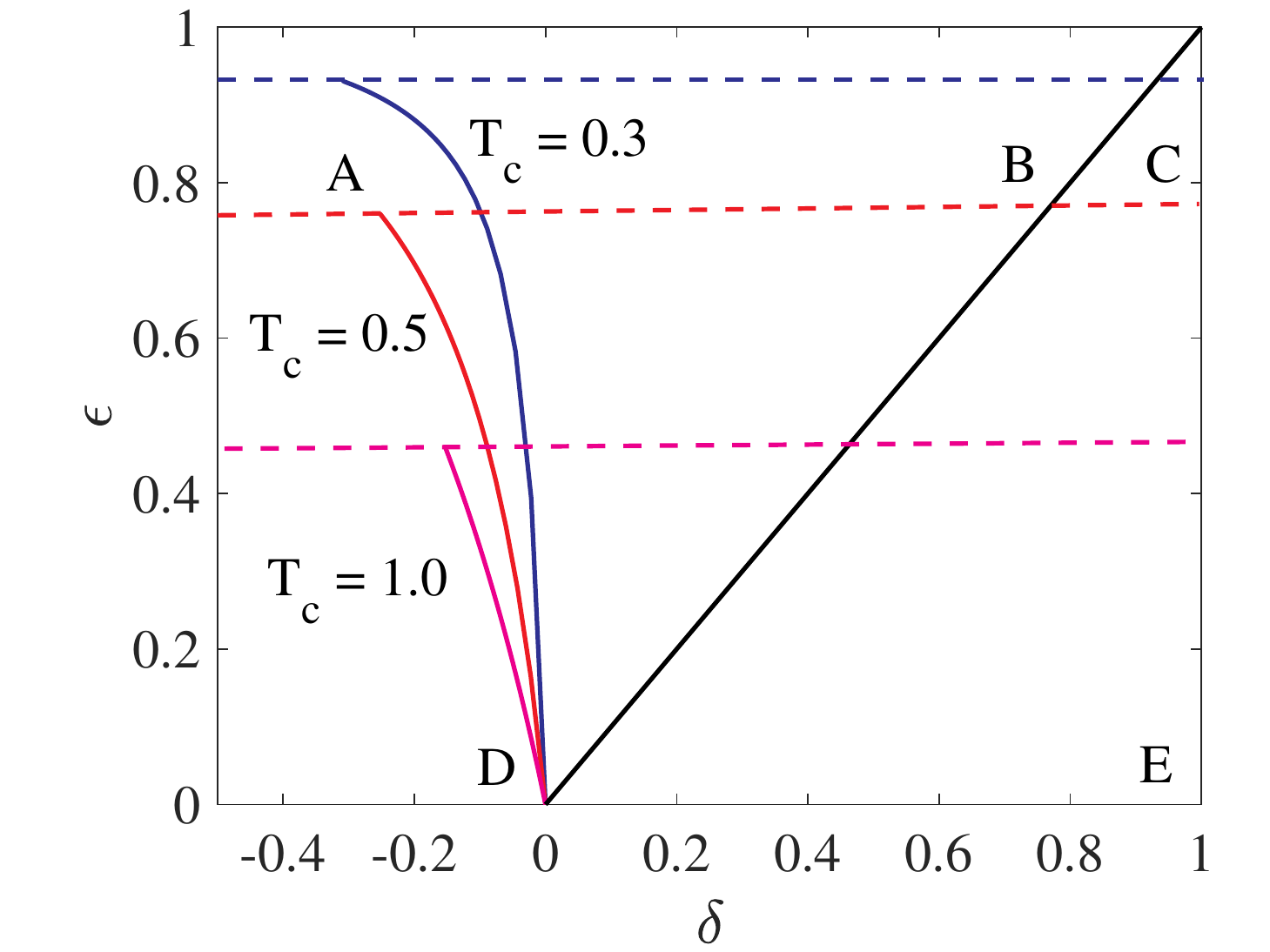}
\caption{The phase diagram of the modified Maxwell refrigerator model in the $\epsilon$ - $\delta$ plane 
for three different values of $T_c$: $T_c = 0.3, 0.5$ and $1.0$. The dashed lines indicate the maximum 
values of $\epsilon$ possible for each of the $T_c$ values. The phase diagram has three regions: 
(i) The far right region ($DBCE$ for $T_c = 0.5$) where the demon and bits together work as a refrigerator. 
In the intermediate region ($ABD$ for $T_c = 0.5$), it acts as an eraser. In the other regions it acts neither 
as a refrigerator nor as an eraser.}
\label{fig.2}
\end{center}
\end{figure}

If $T_h \gg 1$, $\sigma$ can be put approximately to $0$ and we have, $\epsilon \approx \omega$
and $\Phi \approx \frac{\delta - \omega}{4}$. Using these values in Eq. (\ref{sbit}) for the change in
bit entropy, one can solve for $\Delta S_B = 0$ giving the nontrivial root, $\delta_a = -\frac{\omega}{3}$ 
(see supplementary for details) in this limit. $\delta = \epsilon$ is in any case a root even without the above 
approximation. It is difficult to analytically solve $\Delta S_B = 0$ for arbitrary values of $T_c$ 
and $T_h$ but the phase diagram for the model can be found numerically. The phase diagram 
indicating three modes of operation for the demon and bits system as a function of $\delta$ and 
$\epsilon$ values for various fixed values of $T_c$ is shown in Fig.~\ref{fig.2}. 

\begin{figure}
\begin{center}
\includegraphics[height = 6.5cm]{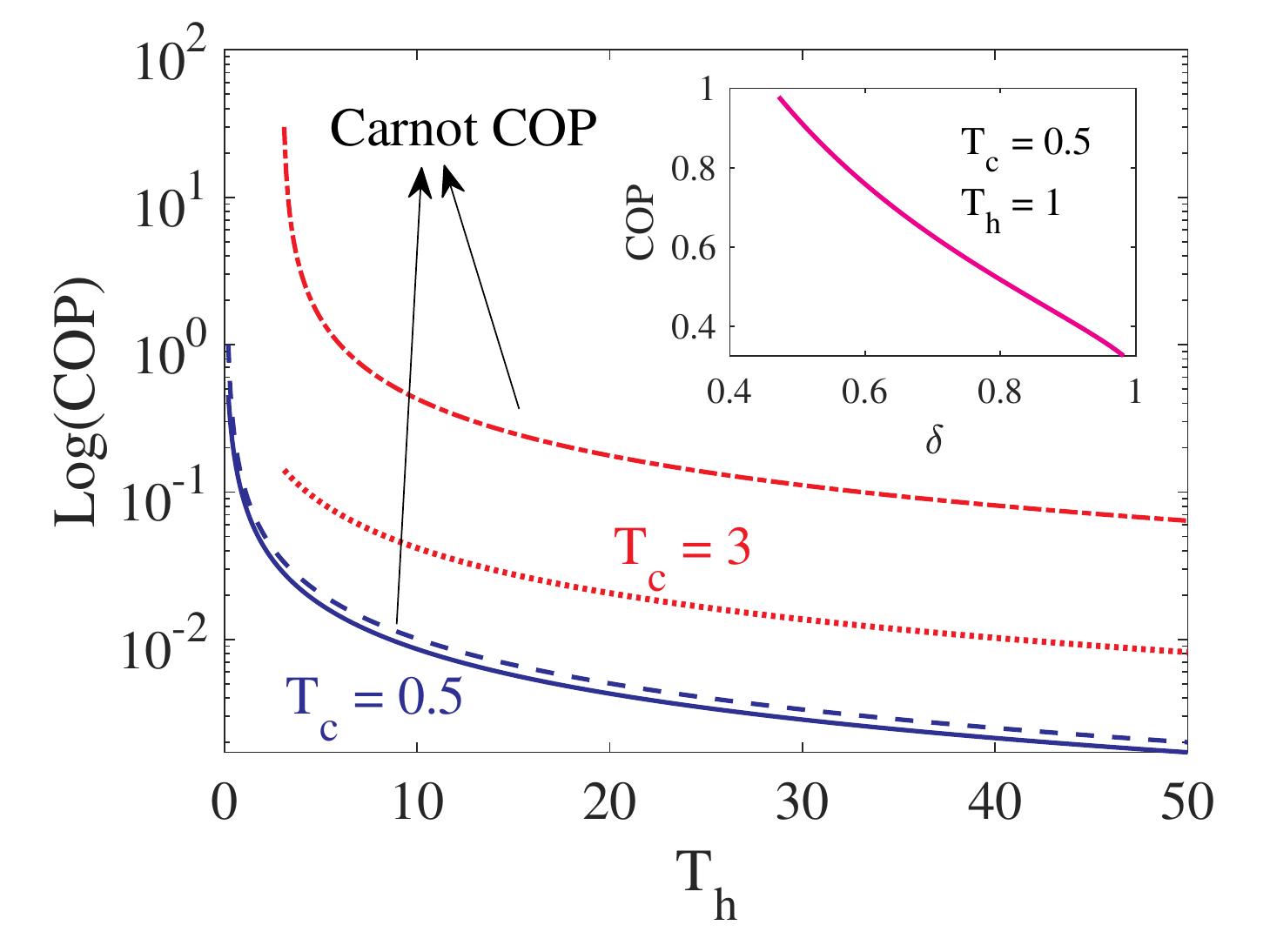}
\caption{The main plot shows the COP at maximum power ($\delta = 1$) for $T_c = 0.5$ (blue line) and 
$T_c = 3$ (red dotted line) as a function of $T_h$. The blue dashed curve is the COP for the Carnot refrigerator 
with $T_c = 0.5$ and the red dash-dot line is the COP for a Carnot refrigerator for $T_c = 3$. The logarithm of COP 
is used to capture the full variation in the data. The inset shows the variation of COP as a function of $\delta$ for 
$T_c = 0.5$ and $T_h = 1$.}
\label{fig.3}
\end{center}
\end{figure}

We now focus on the COP/efficiency of the device in its various regimes of operation. As discussed above when 
$\epsilon < \delta \le 1$, the device acts as a refrigerator transferring heat $\Phi$ from the cold to the hot 
reservoir. For this device to work in a cyclic manner, one needs to reset the outgoing bits, such that the 
probability distribution for $1$s and $0$s are the same as that of the incoming bits. Since $\Delta S_B$ is positive 
in this regime, the resetting process will need a minimum work to be carried out by an amount equal to 
$T_h \Delta S_B$ by Landauer's principle. Note that the resetting has to be done with the bits in contact with 
the hot bath as dissipating heat into the cold bath will be at cross purpose with the intention of cooling that bath. 
Thus the coefficient of performance (COP) of the device when working as a refrigerator is given by,
\begin{equation}
{\rm COP} =  \frac{\Phi}{T_h \Delta S_B} \;.
\label{cop}
\end{equation}
The variation of COP for the case when $T_c = 0.5$ and $T_h = 1$ is given in the inset of Fig.~\ref{fig.3}, 
where we have used the derived forms of $\Phi$ and $\Delta S_B$ (Eqs. (\ref{phi}) and (\ref{sbit}) ) to compute 
the curve. Note that the smallest value of delta in this plot is $\epsilon = 0.46$.  

There are two important limits to look at in the refrigerator operation: (i) $\delta = \epsilon$, where the process is 
reversible and (ii) $\delta = 1$, where $\Phi$ is maximum and hence one gets maximum power. In the 
limit $\delta$ approaches $\epsilon$ from above, one can show that the COP tends to the limit 
$\frac{1}{\frac{T_h}{T_c} - 1}$ (see supplementary). This follows from the observation that 
$-\Delta S_{res}$ is tangential to the $\Delta S_B$ curve at $\delta = \epsilon$ in Fig.~\ref{fig.1}. 
The result is not surprising since in the limit $\delta$ approaches $\epsilon$, the refrigerator cycle 
becomes reversible and one should expect the COP to match with that of the Carnot refrigerator. 
In fact, it is an indication of the fact that our definition of COP is
consistent with its traditional definition. Operating a refrigerator or engine reversibly is not practically very 
interesting as it takes infinite time to complete the cycle \cite{Curzon1975}. Finite time 
thermodynamics looks for optimization also in terms of power or rate at which heat is transferred. In the 
current device working as a refrigerator, maximum power can be attained by maximizing the value of $Q$. 
This happens at $\delta = 1$ where $\Phi$ has the largest positive value. With $\delta = 1$, the change in 
entropy of bits is given by
\begin{equation}
(\Delta S_B)_m = -(1 - \Phi_m) \ln(1 - \Phi_m) - \Phi_m \ln (\Phi_m)
\label{copm}
\end{equation}
where $\Phi_m = \frac{(1 - \epsilon)}{4} (1 - \sigma \omega)$ is $\Phi$ evaluated for $\delta = 1$.
Fig.~\ref{fig.3} shows the COP of the refrigerator at maximum power (that is when $\delta = 1$)
as a function of $T_h$ for two different values of $T_c$: $T_c = 0.5$ and $T_c = 3$. As expected, the 
COP at maximum power lies below the Carnot refrigerator COP. It can be shown that in the limit of $T_h$ and 
$T_c$ both much larger than $1$ (see supplementary),
\begin{equation}
{\rm COP}_{max} = \frac{1}{(4 \ln 4 - 3 \ln 3) T_h}
\label{copml}
\end{equation}
and is independent of $T_c$. 

\begin{figure}
\begin{center}
\includegraphics[height = 6.5cm]{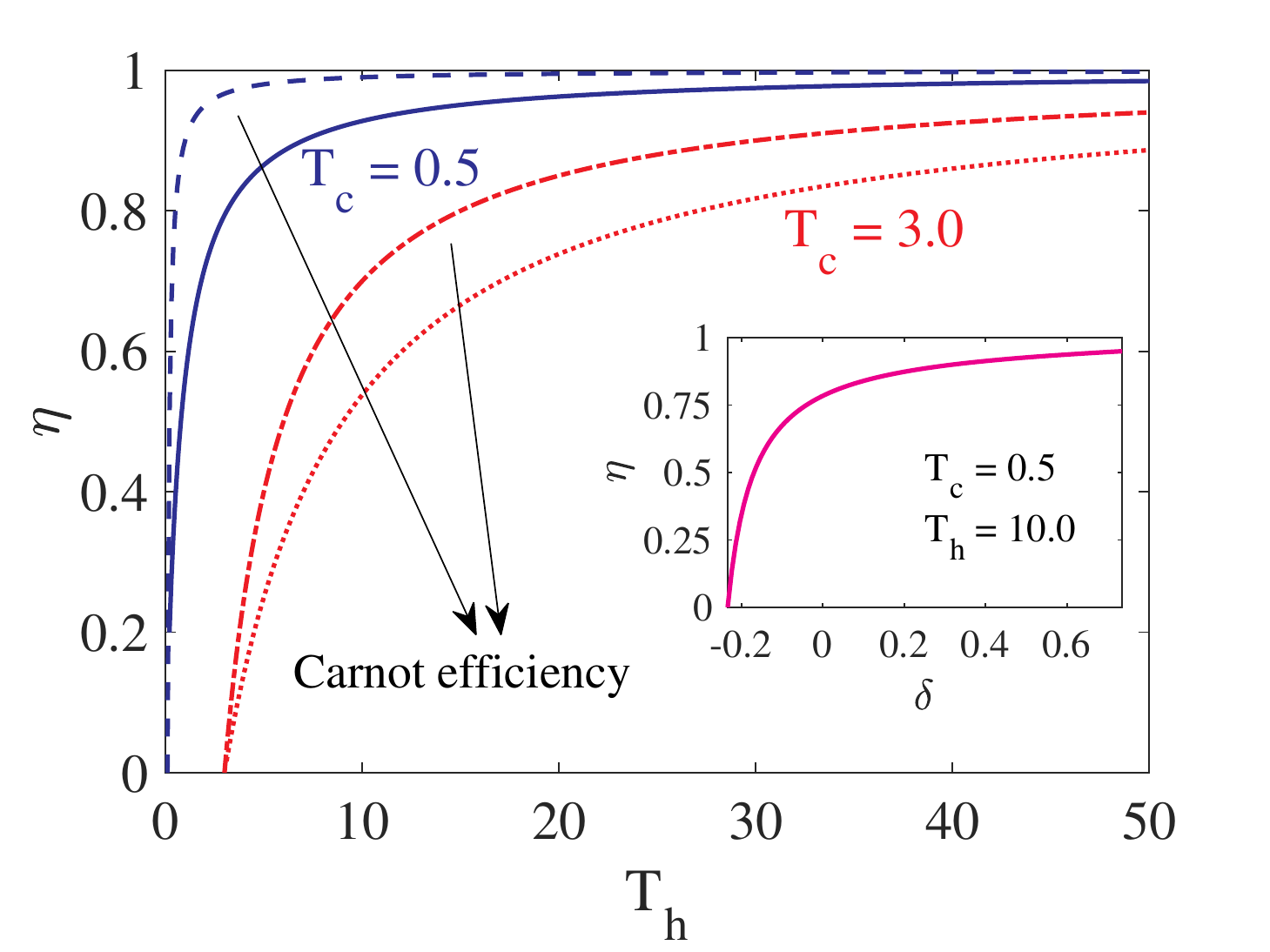}
\caption{The main plot shows the the efficiency of the heat engine at maximum power ($\delta = \delta_m$) 
for $T_c = 0.5$ (blue line) and $T_c = 3$ (red dotted line) as a function of $T_h$. The blue dashed curve is the 
efficiency for the Carnot engine with $T_c = 0.5$ and the red dash-dotted line is the efficiency for a Carnot engine 
for $T_c = 3$. The inset shows the variation of efficiency as a function of $\delta$ for $T_c = 0.5$ and 
$T_h = 10$.}
\label{fig.4}
\end{center}
\end{figure}
We now look at the erasure/heat engine regime of the device. We shall assume extraction of work equal 
$T_h \Delta S_B$ via a reversible process of the kind carried out in a Szilard engine. The work is extracted in 
contact with the hot reservoir as that would give us maximum possible work output for a given $\Delta S_B$. 
The efficiency of the heat engine can be computed as,
\begin{equation}
\eta =  \frac{-T_h \Delta S_B}{-\Phi - T_h \Delta S_B} \;.
\label{eta}
\end{equation}
where the numerator on the right is the work extracted and the denominator is the net heat absorbed from 
the hot reservoir. $\Phi$ and $\Delta S_B$ are negative quantities when $\delta_a < \delta < \epsilon$.  

As in the case of refrigerator COP, one can show that $\eta$ approaches the Carnot efficiency as 
$\delta$ approaches $\epsilon$ from below. This is not surprising as the engine is reversible in that limit. 
The variation of $\eta$ for fixed values of $T_c$ and $T_h$ ($T_c = 0.5$ and $T_h = 10$)are shown in the 
the inset of  Fig.~\ref{fig.4}. The efficiency at maximum power at fixed $T_h$ and $T_c$ can be computed by 
finding the value for $\delta$ for which $\Delta S_B$ is a minimum and evaluating $\eta$ at that value. 
Fig.~\ref{fig.4} shows the variation of efficiency at maximum power as a function of $T_h$ for two fixed 
values of $T_c$. This curve has been computed numerically by finding the root of the equation 
$\frac{d}{d \delta} \Delta S_B(\delta) = 0$ and evaluating the efficiency at that value of $\delta$.
As expected, the efficiency lies below the corresponding Carnot efficiency. It is clear from Eq. (\ref{eta})
that for large values of $T_h$, the efficiency will approach $1$, just as it is for the Carnot engine. 

One can analytically find $\eta_{max}$ for the case when $T_h \gg 1$. The value of $\delta$ leading to the 
minima of $\Delta S_B$ in this limit can be found to be the solution of the cubic equation 
$\delta^3 + \omega \delta^2 - 3 \delta + \omega = 0$(see supplementary) . For this case, if $T_c$ is also 
much larger than one, the region $\delta_a < \delta < \epsilon$ narrows down to zero as both $\delta_a$ 
and $\epsilon$ are proportional to $\omega$ when $T_h$ tends to large values. In the limit $T_c \ll 1$, 
$\omega$ can be approximated to $1$ giving the root as $\delta = \sqrt{2} - 1$. Evaluating the value of 
$\eta$ for this limit one gets,
\begin{equation}
\eta_{max} = \frac{T_h}{0.779 + T_h}
\label{etalt}
\end{equation}
that depends only on $T_h$ and is valid in the limit $T_h \gg 1$ and $T_c \ll 1$.

Efficiencies of engines and refrigerators at maximum power has been evaluated for a number of 
systems both macroscopic \cite{NOVIKOV1958125,Curzon1975,PhysRevLett.78.3241} and microscopic 
\cite{Schmiedl_2007,Tu_2008}. The efficiency at maximum power depends on the particular model 
as well as the way optimization is carried out \cite{Seifert2012}. In the linear regime, with strong coupling 
and left-right symmetry, it has been shown that Curzon-Ahlborn result is universal up to second order in 
$\eta_C$, the Carnot efficiency \cite{PhysRevLett.102.130602}. Note that the results presented in this work
does not depend on the $T_h \approx T_c$ limit but is an exact analysis of the particular model system studied.
The presence of the energy scale corresponding to working substance (the level spacing of the demon) of
the microscopic device is what allows for the various approximations we have carried out.

It is to be kept in mind that the power optimization was done without taking into consideration the 
power maximization related to work done on resetting the bit (for refrigerator mode) or work extraction from 
erased bits (for the engine mode). As far as the COP  of the refrigerator is concerned this will lead to a 
multiplication by a constant factor less than one because at maximum power the resetting will
take more work than $T_h \Delta S_B$ to be performed. For the engine efficiency, power maximization in
the work extraction process using the low entropy outgoing bit stream will lead to a value of work obtained 
that is less in magnitude than $T_h \Delta S_B$ in Eq. (\ref{eta}). This would imply that the $\eta_{max}$ we 
found will just be an upper bound for the maximum power for the full device.

To summarize, we have studied a modified version of Maxwell refrigerator in which the bit demon system
equilibrate with the thermal reservoirs alternately. The analysis of the current version is simple and allows one
to study the efficiency of the device in detail. The second law of thermodynamics has been shown to hold
in its generalized version which includes the Shannon entropy of the bit stream. The phase diagram of the
device shows that the device can work as a refrigerator as well as a heat engine much like the original
Maxwell refrigerator \cite{Mandal2013a}. The power of the device has been maximized using the probability
distribution of the incoming bits as the variational parameter. The efficiency studies lead to the interesting results 
that both the COP of the refrigerator and efficiency of the heat engine at maximum power can be independent of
the temperature of the low temperature bath in the appropriate limits. These results could be of relevance
to cellular level biological process, design of artificial molecular machines and thermodynamics of computation.
The work could be generalized by making the interaction time of the device with the baths to be finite rather than
letting it equilibrate. Another avenue to be explored will be quantum version of the device along the lines
of recent work on a heat engine with two level system as the working substance \cite{Erdman_2019}.

%\bibliography{Paper-arxiv}

\begin{thebibliography}{34}%
\makeatletter
\providecommand \@ifxundefined [1]{%
 \@ifx{#1\undefined}
}%
\providecommand \@ifnum [1]{%
 \ifnum #1\expandafter \@firstoftwo
 \else \expandafter \@secondoftwo
 \fi
}%
\providecommand \@ifx [1]{%
 \ifx #1\expandafter \@firstoftwo
 \else \expandafter \@secondoftwo
 \fi
}%
\providecommand \natexlab [1]{#1}%
\providecommand \enquote  [1]{``#1''}%
\providecommand \bibnamefont  [1]{#1}%
\providecommand \bibfnamefont [1]{#1}%
\providecommand \citenamefont [1]{#1}%
\providecommand \href@noop [0]{\@secondoftwo}%
\providecommand \href [0]{\begingroup \@sanitize@url \@href}%
\providecommand \@href[1]{\@@startlink{#1}\@@href}%
\providecommand \@@href[1]{\endgroup#1\@@endlink}%
\providecommand \@sanitize@url [0]{\catcode `\\12\catcode `\$12\catcode
  `\&12\catcode `\#12\catcode `\^12\catcode `\_12\catcode `\%12\relax}%
\providecommand \@@startlink[1]{}%
\providecommand \@@endlink[0]{}%
\providecommand \url  [0]{\begingroup\@sanitize@url \@url }%
\providecommand \@url [1]{\endgroup\@href {#1}{\urlprefix }}%
\providecommand \urlprefix  [0]{URL }%
\providecommand \Eprint [0]{\href }%
\providecommand \doibase [0]{http://dx.doi.org/}%
\providecommand \selectlanguage [0]{\@gobble}%
\providecommand \bibinfo  [0]{\@secondoftwo}%
\providecommand \bibfield  [0]{\@secondoftwo}%
\providecommand \translation [1]{[#1]}%
\providecommand \BibitemOpen [0]{}%
\providecommand \bibitemStop [0]{}%
\providecommand \bibitemNoStop [0]{.\EOS\space}%
\providecommand \EOS [0]{\spacefactor3000\relax}%
\providecommand \BibitemShut  [1]{\csname bibitem#1\endcsname}%
\let\auto@bib@innerbib\@empty
%</preamble>
\bibitem [{\citenamefont {Mandal}\ \emph {et~al.}(2013)\citenamefont {Mandal},
  \citenamefont {Quan},\ and\ \citenamefont {Jarzynski}}]{Mandal2013a}%
  \BibitemOpen
  \bibfield  {author} {\bibinfo {author} {\bibfnamefont {D.}~\bibnamefont
  {Mandal}}, \bibinfo {author} {\bibfnamefont {H.~T.}\ \bibnamefont {Quan}}, \
  and\ \bibinfo {author} {\bibfnamefont {C.}~\bibnamefont {Jarzynski}},\ }\href
  {\doibase 10.1103/PhysRevLett.111.030602} {\bibfield  {journal} {\bibinfo
  {journal} {Phys. Rev. Lett.}\ }\textbf {\bibinfo {volume} {111}},\ \bibinfo
  {pages} {030602} (\bibinfo {year} {2013})}\BibitemShut {NoStop}%
\bibitem [{\citenamefont {{Clerk Maxwell}}(1871)}]{ClerkMaxwell:1871:TH}%
  \BibitemOpen
  \bibfield  {author} {\bibinfo {author} {\bibfnamefont {J.}~\bibnamefont
  {{Clerk Maxwell}}},\ }\href {https://hdl.handle.net/2027/nyp.33433057781498;
  https://hdl.handle.net/2027/nyp.33433069099806} {\emph {\bibinfo {title}
  {Theory of Heat}}}\ (\bibinfo  {publisher} {Longmans, Green, and Co.},\
  \bibinfo {address} {London, UK},\ \bibinfo {year} {1871})\BibitemShut
  {NoStop}%
\bibitem [{\citenamefont {M.}(1912)}]{Smoluchowski1912}%
  \BibitemOpen
  \bibfield  {author} {\bibinfo {author} {\bibfnamefont {S.}~\bibnamefont
  {M.}},\ }\href@noop {} {\bibfield  {journal} {\bibinfo  {journal} {Phys. Z.}\
  }\textbf {\bibinfo {volume} {13}},\ \bibinfo {pages} {1069} (\bibinfo {year}
  {1912})}\BibitemShut {NoStop}%
\bibitem [{\citenamefont {Szilard}(1929)}]{Szilard:1929:EET}%
  \BibitemOpen
  \bibfield  {author} {\bibinfo {author} {\bibfnamefont {L.}~\bibnamefont
  {Szilard}},\ }\href {\doibase https://doi.org/10.1007/BF01341281} {\ \textbf
  {\bibinfo {volume} {53}},\ \bibinfo {pages} {840} (\bibinfo {year}
  {1929})}\BibitemShut {NoStop}%
\bibitem [{\citenamefont {Feynman}\ \emph {et~al.}(1965)\citenamefont
  {Feynman}, \citenamefont {Leighton},\ and\ \citenamefont
  {Sands}}]{Feynman:1963:FLP}%
  \BibitemOpen
  \bibfield  {author} {\bibinfo {author} {\bibfnamefont {R.~P. R.~P.}\
  \bibnamefont {Feynman}}, \bibinfo {author} {\bibfnamefont {R.~B.}\
  \bibnamefont {Leighton}}, \ and\ \bibinfo {author} {\bibfnamefont {M.~L.
  M.~L.}\ \bibnamefont {Sands}},\ }\href@noop {} {\emph {\bibinfo {title} {The
  {Feynman} lectures on physics - Vol 1}}}\ (\bibinfo {year}
  {1965})\BibitemShut {NoStop}%
\bibitem [{\citenamefont {Brillouin}(1951)}]{Brillouin1951}%
  \BibitemOpen
  \bibfield  {author} {\bibinfo {author} {\bibfnamefont {L.}~\bibnamefont
  {Brillouin}},\ }\href {\doibase 10.1063/1.1699951} {\bibfield  {journal}
  {\bibinfo  {journal} {Journal of Applied Physics}\ }\textbf {\bibinfo
  {volume} {22}},\ \bibinfo {pages} {334} (\bibinfo {year} {1951})}\BibitemShut
  {NoStop}%
\bibitem [{\citenamefont {Penrose}(1937)}]{OPenrose1979}%
  \BibitemOpen
  \bibfield  {author} {\bibinfo {author} {\bibfnamefont {O.}~\bibnamefont
  {Penrose}},\ }\href@noop {} {\bibfield  {journal} {\bibinfo  {journal} {Rep.
  Prog. Phys.}\ }\textbf {\bibinfo {volume} {42}},\ \bibinfo {pages} {1938}
  (\bibinfo {year} {1937})}\BibitemShut {NoStop}%
\bibitem [{\citenamefont {Landauer}(1961)}]{Landauer1961}%
  \BibitemOpen
  \bibfield  {author} {\bibinfo {author} {\bibfnamefont {R.}~\bibnamefont
  {Landauer}},\ }\href@noop {} {\bibfield  {journal} {\bibinfo  {journal} {IBM
  J. Res. Dev.}\ }\textbf {\bibinfo {volume} {5}},\ \bibinfo {pages} {183}
  (\bibinfo {year} {1961})}\BibitemShut {NoStop}%
\bibitem [{\citenamefont {Bennett}(1982)}]{Bennett1982}%
  \BibitemOpen
  \bibfield  {author} {\bibinfo {author} {\bibfnamefont {C.~H.}\ \bibnamefont
  {Bennett}},\ }\href@noop {} {\bibfield  {journal} {\bibinfo  {journal} {Int.
  J. Theor. Phys.}\ }\textbf {\bibinfo {volume} {21}},\ \bibinfo {pages} {905}
  (\bibinfo {year} {1982})}\BibitemShut {NoStop}%
\bibitem [{\citenamefont {Bennett}\ and\ \citenamefont
  {Landauer}(1985)}]{Bennett1985}%
  \BibitemOpen
  \bibfield  {author} {\bibinfo {author} {\bibfnamefont {C.~H.}\ \bibnamefont
  {Bennett}}\ and\ \bibinfo {author} {\bibfnamefont {R.}~\bibnamefont
  {Landauer}},\ }\href {\doibase 10.1038/scientificamerican0785-48} {\bibfield
  {journal} {\bibinfo  {journal} {Sci. Am.}\ }\textbf {\bibinfo {volume}
  {253}},\ \bibinfo {pages} {48} (\bibinfo {year} {1985})}\BibitemShut
  {NoStop}%
\bibitem [{\citenamefont {Maruyama}\ \emph {et~al.}(2009)\citenamefont
  {Maruyama}, \citenamefont {Nori},\ and\ \citenamefont
  {Vedral}}]{Maruyama2009}%
  \BibitemOpen
  \bibfield  {author} {\bibinfo {author} {\bibfnamefont {K.}~\bibnamefont
  {Maruyama}}, \bibinfo {author} {\bibfnamefont {F.}~\bibnamefont {Nori}}, \
  and\ \bibinfo {author} {\bibfnamefont {V.}~\bibnamefont {Vedral}},\ }\href
  {\doibase 10.1103/RevModPhys.81.1} {\bibfield  {journal} {\bibinfo  {journal}
  {Rev. Mod. Phys.}\ }\textbf {\bibinfo {volume} {81}},\ \bibinfo {pages} {1}
  (\bibinfo {year} {2009})}\BibitemShut {NoStop}%
\bibitem [{\citenamefont {B{\'{e}}rut}\ \emph {et~al.}(2012)\citenamefont
  {B{\'{e}}rut}, \citenamefont {Arakelyan}, \citenamefont {Petrosyan},
  \citenamefont {Ciliberto}, \citenamefont {Dillenschneider},\ and\
  \citenamefont {Lutz}}]{Berut2012}%
  \BibitemOpen
  \bibfield  {author} {\bibinfo {author} {\bibfnamefont {A.}~\bibnamefont
  {B{\'{e}}rut}}, \bibinfo {author} {\bibfnamefont {A.}~\bibnamefont
  {Arakelyan}}, \bibinfo {author} {\bibfnamefont {A.}~\bibnamefont
  {Petrosyan}}, \bibinfo {author} {\bibfnamefont {S.}~\bibnamefont
  {Ciliberto}}, \bibinfo {author} {\bibfnamefont {R.}~\bibnamefont
  {Dillenschneider}}, \ and\ \bibinfo {author} {\bibfnamefont {E.}~\bibnamefont
  {Lutz}},\ }\href {\doibase 10.1038/nature10872} {\bibfield  {journal}
  {\bibinfo  {journal} {Nature}\ }\textbf {\bibinfo {volume} {483}},\ \bibinfo
  {pages} {187} (\bibinfo {year} {2012})}\BibitemShut {NoStop}%
\bibitem [{\citenamefont {Toyabe}\ \emph {et~al.}(2010)\citenamefont {Toyabe},
  \citenamefont {Sagawa}, \citenamefont {Ueda}, \citenamefont {Muneyuki},\ and\
  \citenamefont {Sano}}]{Toyabe2010}%
  \BibitemOpen
  \bibfield  {author} {\bibinfo {author} {\bibfnamefont {S.}~\bibnamefont
  {Toyabe}}, \bibinfo {author} {\bibfnamefont {T.}~\bibnamefont {Sagawa}},
  \bibinfo {author} {\bibfnamefont {M.}~\bibnamefont {Ueda}}, \bibinfo {author}
  {\bibfnamefont {E.}~\bibnamefont {Muneyuki}}, \ and\ \bibinfo {author}
  {\bibfnamefont {M.}~\bibnamefont {Sano}},\ }\href@noop {} {\bibfield
  {journal} {\bibinfo  {journal} {Nat. Phys.}\ }\textbf {\bibinfo {volume}
  {6}},\ \bibinfo {pages} {988} (\bibinfo {year} {2010})}\BibitemShut {NoStop}%
\bibitem [{\citenamefont {Masuyama}\ \emph {et~al.}(2018)\citenamefont
  {Masuyama}, \citenamefont {Funo}, \citenamefont {Murashita}, \citenamefont
  {Noguchi}, \citenamefont {Kono}, \citenamefont {Tabuchi}, \citenamefont
  {Yamazaki}, \citenamefont {Ueda},\ and\ \citenamefont
  {Nakamura}}]{Masuyama2018}%
  \BibitemOpen
  \bibfield  {author} {\bibinfo {author} {\bibfnamefont {Y.}~\bibnamefont
  {Masuyama}}, \bibinfo {author} {\bibfnamefont {K.}~\bibnamefont {Funo}},
  \bibinfo {author} {\bibfnamefont {Y.}~\bibnamefont {Murashita}}, \bibinfo
  {author} {\bibfnamefont {A.}~\bibnamefont {Noguchi}}, \bibinfo {author}
  {\bibfnamefont {S.}~\bibnamefont {Kono}}, \bibinfo {author} {\bibfnamefont
  {Y.}~\bibnamefont {Tabuchi}}, \bibinfo {author} {\bibfnamefont
  {R.}~\bibnamefont {Yamazaki}}, \bibinfo {author} {\bibfnamefont
  {M.}~\bibnamefont {Ueda}}, \ and\ \bibinfo {author} {\bibfnamefont
  {Y.}~\bibnamefont {Nakamura}},\ }\href {\doibase 10.1038/s41467-018-03686-y}
  {\bibfield  {journal} {\bibinfo  {journal} {Nature Communications}\ }\textbf
  {\bibinfo {volume} {9}},\ \bibinfo {pages} {1291} (\bibinfo {year}
  {2018})}\BibitemShut {NoStop}%
\bibitem [{\citenamefont {Paneru}\ \emph {et~al.}(2020)\citenamefont {Paneru},
  \citenamefont {Dutta}, \citenamefont {Sagawa}, \citenamefont {Tlusty},\ and\
  \citenamefont {Pak}}]{Paneru2020}%
  \BibitemOpen
  \bibfield  {author} {\bibinfo {author} {\bibfnamefont {G.}~\bibnamefont
  {Paneru}}, \bibinfo {author} {\bibfnamefont {S.}~\bibnamefont {Dutta}},
  \bibinfo {author} {\bibfnamefont {T.}~\bibnamefont {Sagawa}}, \bibinfo
  {author} {\bibfnamefont {T.}~\bibnamefont {Tlusty}}, \ and\ \bibinfo {author}
  {\bibfnamefont {H.~K.}\ \bibnamefont {Pak}},\ }\href {\doibase
  10.1038/s41467-020-14823-x} {\bibfield  {journal} {\bibinfo  {journal}
  {Nature Communications}\ }\textbf {\bibinfo {volume} {11}},\ \bibinfo {pages}
  {1012} (\bibinfo {year} {2020})}\BibitemShut {NoStop}%
\bibitem [{\citenamefont {Ito}\ and\ \citenamefont {Sagawa}(2015)}]{Ito2015}%
  \BibitemOpen
  \bibfield  {author} {\bibinfo {author} {\bibfnamefont {S.}~\bibnamefont
  {Ito}}\ and\ \bibinfo {author} {\bibfnamefont {T.}~\bibnamefont {Sagawa}},\
  }\href {\doibase 10.1038/ncomms8498} {\bibfield  {journal} {\bibinfo
  {journal} {Nature Communications}\ }\textbf {\bibinfo {volume} {6}},\
  \bibinfo {pages} {7498} (\bibinfo {year} {2015})}\BibitemShut {NoStop}%
\bibitem [{\citenamefont {Tu}(2008{\natexlab{a}})}]{Tu11737}%
  \BibitemOpen
  \bibfield  {author} {\bibinfo {author} {\bibfnamefont {Y.}~\bibnamefont
  {Tu}},\ }\href {\doibase 10.1073/pnas.0804641105} {\bibfield  {journal}
  {\bibinfo  {journal} {Proceedings of the National Academy of Sciences}\
  }\textbf {\bibinfo {volume} {105}},\ \bibinfo {pages} {11737} (\bibinfo
  {year} {2008}{\natexlab{a}})}\BibitemShut {NoStop}%
\bibitem [{\citenamefont {Kay}\ and\ \citenamefont {Leigh}(2015)}]{Euan2015}%
  \BibitemOpen
  \bibfield  {author} {\bibinfo {author} {\bibfnamefont {E.~R.}\ \bibnamefont
  {Kay}}\ and\ \bibinfo {author} {\bibfnamefont {D.~A.}\ \bibnamefont
  {Leigh}},\ }\href {\doibase 10.1002/anie.201503375} {\bibfield  {journal}
  {\bibinfo  {journal} {Angewandte Chemie International Edition}\ }\textbf
  {\bibinfo {volume} {54}},\ \bibinfo {pages} {10080} (\bibinfo {year}
  {2015})}\BibitemShut {NoStop}%
\bibitem [{\citenamefont {Quan}\ \emph {et~al.}(2006)\citenamefont {Quan},
  \citenamefont {Wang}, \citenamefont {Liu}, \citenamefont {Sun},\ and\
  \citenamefont {Nori}}]{Quan2006}%
  \BibitemOpen
  \bibfield  {author} {\bibinfo {author} {\bibfnamefont {H.~T.}\ \bibnamefont
  {Quan}}, \bibinfo {author} {\bibfnamefont {Y.~D.}\ \bibnamefont {Wang}},
  \bibinfo {author} {\bibfnamefont {Y.-x.}\ \bibnamefont {Liu}}, \bibinfo
  {author} {\bibfnamefont {C.~P.}\ \bibnamefont {Sun}}, \ and\ \bibinfo
  {author} {\bibfnamefont {F.}~\bibnamefont {Nori}},\ }\href {\doibase
  10.1103/PhysRevLett.97.180402} {\bibfield  {journal} {\bibinfo  {journal}
  {Phys. Rev. Lett.}\ }\textbf {\bibinfo {volume} {97}},\ \bibinfo {pages}
  {180402} (\bibinfo {year} {2006})}\BibitemShut {NoStop}%
\bibitem [{\citenamefont {Abreu}\ and\ \citenamefont
  {Seifert}(2011)}]{Abreu2011}%
  \BibitemOpen
  \bibfield  {author} {\bibinfo {author} {\bibfnamefont {D.}~\bibnamefont
  {Abreu}}\ and\ \bibinfo {author} {\bibfnamefont {U.}~\bibnamefont
  {Seifert}},\ }\href {\doibase 10 001} {\bibfield  {journal} {\bibinfo
  {journal} {Epl}\ }\textbf {\bibinfo {volume} {94}} (\bibinfo {year} {2011}),\
  10 001}\BibitemShut {NoStop}%
\bibitem [{\citenamefont {Mandal}\ and\ \citenamefont
  {Jarzynski}(2012)}]{Mandal2012}%
  \BibitemOpen
  \bibfield  {author} {\bibinfo {author} {\bibfnamefont {D.}~\bibnamefont
  {Mandal}}\ and\ \bibinfo {author} {\bibfnamefont {C.}~\bibnamefont
  {Jarzynski}},\ }\href {\doibase 10.1073/pnas.1204263109} {\bibfield
  {journal} {\bibinfo  {journal} {Proceedings of the National Academy of
  Sciences}\ }\textbf {\bibinfo {volume} {109}},\ \bibinfo {pages} {11641}
  (\bibinfo {year} {2012})}\BibitemShut {NoStop}%
\bibitem [{\citenamefont {Barato}\ and\ \citenamefont
  {Seifert}(2014)}]{Barato2014a}%
  \BibitemOpen
  \bibfield  {author} {\bibinfo {author} {\bibfnamefont {A.~C.}\ \bibnamefont
  {Barato}}\ and\ \bibinfo {author} {\bibfnamefont {U.}~\bibnamefont
  {Seifert}},\ }\href {\doibase 10.1103/PhysRevLett.112.090601} {\bibfield
  {journal} {\bibinfo  {journal} {Phys. Rev. Lett.}\ }\textbf {\bibinfo
  {volume} {112}},\ \bibinfo {pages} {090601} (\bibinfo {year}
  {2014})}\BibitemShut {NoStop}%
\bibitem [{\citenamefont {Ribezzi-Crivellari}\ and\ \citenamefont
  {Ritort}(2019)}]{Ribezzi-Crivellari2019}%
  \BibitemOpen
  \bibfield  {author} {\bibinfo {author} {\bibfnamefont {M.}~\bibnamefont
  {Ribezzi-Crivellari}}\ and\ \bibinfo {author} {\bibfnamefont
  {F.}~\bibnamefont {Ritort}},\ }\href {\doibase 10.1038/s41567-019-0481-0}
  {\bibfield  {journal} {\bibinfo  {journal} {Nature Physics}\ }\textbf
  {\bibinfo {volume} {15}},\ \bibinfo {pages} {660} (\bibinfo {year}
  {2019})}\BibitemShut {NoStop}%
\bibitem [{\citenamefont {Seifert}(2012)}]{Seifert2012}%
  \BibitemOpen
  \bibfield  {author} {\bibinfo {author} {\bibfnamefont {U.}~\bibnamefont
  {Seifert}},\ }\href {\doibase 10.1088/0034-4885/75/12/126001} {\bibfield
  {journal} {\bibinfo  {journal} {Reports Prog. Phys.}\ }\textbf {\bibinfo
  {volume} {75}},\ \bibinfo {pages} {126001} (\bibinfo {year}
  {2012})}\BibitemShut {NoStop}%
\bibitem [{\citenamefont {Kim}\ and\ \citenamefont {Qian}(2007)}]{Kim2007}%
  \BibitemOpen
  \bibfield  {author} {\bibinfo {author} {\bibfnamefont {K.~H.}\ \bibnamefont
  {Kim}}\ and\ \bibinfo {author} {\bibfnamefont {H.}~\bibnamefont {Qian}},\
  }\href {\doibase 10.1103/PhysRevE.75.022102} {\bibfield  {journal} {\bibinfo
  {journal} {Phys. Rev. E}\ }\textbf {\bibinfo {volume} {75}},\ \bibinfo
  {pages} {022102} (\bibinfo {year} {2007})}\BibitemShut {NoStop}%
\bibitem [{\citenamefont {Sagawa}\ and\ \citenamefont
  {Ueda}(2008)}]{Sagawa2008}%
  \BibitemOpen
  \bibfield  {author} {\bibinfo {author} {\bibfnamefont {T.}~\bibnamefont
  {Sagawa}}\ and\ \bibinfo {author} {\bibfnamefont {M.}~\bibnamefont {Ueda}},\
  }\href {\doibase 10.1103/PhysRevLett.100.080403} {\bibfield  {journal}
  {\bibinfo  {journal} {Phys. Rev. Lett.}\ }\textbf {\bibinfo {volume} {100}},\
  \bibinfo {pages} {080403} (\bibinfo {year} {2008})}\BibitemShut {NoStop}%
\bibitem [{\citenamefont {Annby-Andersson}\ \emph {et~al.}(2020)\citenamefont
  {Annby-Andersson}, \citenamefont {Samuelsson}, \citenamefont {Maisi},\ and\
  \citenamefont {Potts}}]{Annby2020}%
  \BibitemOpen
  \bibfield  {author} {\bibinfo {author} {\bibfnamefont {B.}~\bibnamefont
  {Annby-Andersson}}, \bibinfo {author} {\bibfnamefont {P.}~\bibnamefont
  {Samuelsson}}, \bibinfo {author} {\bibfnamefont {V.~F.}\ \bibnamefont
  {Maisi}}, \ and\ \bibinfo {author} {\bibfnamefont {P.~P.}\ \bibnamefont
  {Potts}},\ }\href {\doibase 10.1103/PhysRevB.101.165404} {\bibfield
  {journal} {\bibinfo  {journal} {Phys. Rev. B}\ }\textbf {\bibinfo {volume}
  {101}},\ \bibinfo {pages} {165404} (\bibinfo {year} {2020})}\BibitemShut
  {NoStop}%
\bibitem [{\citenamefont {Curzon}\ and\ \citenamefont
  {Ahlborn}(1975)}]{Curzon1975}%
  \BibitemOpen
  \bibfield  {author} {\bibinfo {author} {\bibfnamefont {F.~L.}\ \bibnamefont
  {Curzon}}\ and\ \bibinfo {author} {\bibfnamefont {B.}~\bibnamefont
  {Ahlborn}},\ }\href {\doibase 10.1119/1.10023} {\bibfield  {journal}
  {\bibinfo  {journal} {American Journal of Physics}\ }\textbf {\bibinfo
  {volume} {43}},\ \bibinfo {pages} {22} (\bibinfo {year} {1975})},\ \Eprint
  {http://arxiv.org/abs/https://doi.org/10.1119/1.10023}
  {https://doi.org/10.1119/1.10023} \BibitemShut {NoStop}%
\bibitem [{\citenamefont {Novikov}(1958)}]{NOVIKOV1958125}%
  \BibitemOpen
  \bibfield  {author} {\bibinfo {author} {\bibfnamefont {I.}~\bibnamefont
  {Novikov}},\ }\href {\doibase https://doi.org/10.1016/0891-3919(58)90244-4}
  {\bibfield  {journal} {\bibinfo  {journal} {Journal of Nuclear Energy
  (1954)}\ }\textbf {\bibinfo {volume} {7}},\ \bibinfo {pages} {125 } (\bibinfo
  {year} {1958})}\BibitemShut {NoStop}%
\bibitem [{\citenamefont {Velasco}\ \emph {et~al.}(1997)\citenamefont
  {Velasco}, \citenamefont {Roco}, \citenamefont {Medina},\ and\ \citenamefont
  {Hern\'andez}}]{PhysRevLett.78.3241}%
  \BibitemOpen
  \bibfield  {author} {\bibinfo {author} {\bibfnamefont {S.}~\bibnamefont
  {Velasco}}, \bibinfo {author} {\bibfnamefont {J.~M.~M.}\ \bibnamefont
  {Roco}}, \bibinfo {author} {\bibfnamefont {A.}~\bibnamefont {Medina}}, \ and\
  \bibinfo {author} {\bibfnamefont {A.~C.}\ \bibnamefont {Hern\'andez}},\
  }\href {\doibase 10.1103/PhysRevLett.78.3241} {\bibfield  {journal} {\bibinfo
   {journal} {Phys. Rev. Lett.}\ }\textbf {\bibinfo {volume} {78}},\ \bibinfo
  {pages} {3241} (\bibinfo {year} {1997})}\BibitemShut {NoStop}%
\bibitem [{\citenamefont {Schmiedl}\ and\ \citenamefont
  {Seifert}(2007)}]{Schmiedl_2007}%
  \BibitemOpen
  \bibfield  {author} {\bibinfo {author} {\bibfnamefont {T.}~\bibnamefont
  {Schmiedl}}\ and\ \bibinfo {author} {\bibfnamefont {U.}~\bibnamefont
  {Seifert}},\ }\href {\doibase 10.1209/0295-5075/81/20003} {\bibfield
  {journal} {\bibinfo  {journal} {{EPL} (Europhysics Letters)}\ }\textbf
  {\bibinfo {volume} {81}},\ \bibinfo {pages} {20003} (\bibinfo {year}
  {2007})}\BibitemShut {NoStop}%
\bibitem [{\citenamefont {Tu}(2008{\natexlab{b}})}]{Tu_2008}%
  \BibitemOpen
  \bibfield  {author} {\bibinfo {author} {\bibfnamefont {Z.~C.}\ \bibnamefont
  {Tu}},\ }\href {\doibase 10.1088/1751-8113/41/31/312003} {\bibfield
  {journal} {\bibinfo  {journal} {Journal of Physics A: Mathematical and
  Theoretical}\ }\textbf {\bibinfo {volume} {41}},\ \bibinfo {pages} {312003}
  (\bibinfo {year} {2008}{\natexlab{b}})}\BibitemShut {NoStop}%
\bibitem [{\citenamefont {Esposito}\ \emph {et~al.}(2009)\citenamefont
  {Esposito}, \citenamefont {Lindenberg},\ and\ \citenamefont {Van~den
  Broeck}}]{PhysRevLett.102.130602}%
  \BibitemOpen
  \bibfield  {author} {\bibinfo {author} {\bibfnamefont {M.}~\bibnamefont
  {Esposito}}, \bibinfo {author} {\bibfnamefont {K.}~\bibnamefont
  {Lindenberg}}, \ and\ \bibinfo {author} {\bibfnamefont {C.}~\bibnamefont
  {Van~den Broeck}},\ }\href {\doibase 10.1103/PhysRevLett.102.130602}
  {\bibfield  {journal} {\bibinfo  {journal} {Phys. Rev. Lett.}\ }\textbf
  {\bibinfo {volume} {102}},\ \bibinfo {pages} {130602} (\bibinfo {year}
  {2009})}\BibitemShut {NoStop}%
\bibitem [{\citenamefont {Erdman}\ \emph {et~al.}(2019)\citenamefont {Erdman},
  \citenamefont {Cavina}, \citenamefont {Fazio}, \citenamefont {Taddei},\ and\
  \citenamefont {Giovannetti}}]{Erdman_2019}%
  \BibitemOpen
  \bibfield  {author} {\bibinfo {author} {\bibfnamefont {P.~A.}\ \bibnamefont
  {Erdman}}, \bibinfo {author} {\bibfnamefont {V.}~\bibnamefont {Cavina}},
  \bibinfo {author} {\bibfnamefont {R.}~\bibnamefont {Fazio}}, \bibinfo
  {author} {\bibfnamefont {F.}~\bibnamefont {Taddei}}, \ and\ \bibinfo {author}
  {\bibfnamefont {V.}~\bibnamefont {Giovannetti}},\ }\href {\doibase
  10.1088/1367-2630/ab4dca} {\bibfield  {journal} {\bibinfo  {journal} {New
  Journal of Physics}\ }\textbf {\bibinfo {volume} {21}},\ \bibinfo {pages}
  {103049} (\bibinfo {year} {2019})}\BibitemShut {NoStop}%
\end{thebibliography}
%merlin.mbs apsrev4-1.bst 2010-07-25 4.21a (PWD, AO, DPC) hacked
%Control: key (0)
%Control: author (72) initials jnrlst
%Control: editor formatted (1) identically to author
%Control: production of article title (-1) disabled
%Control: page (0) single
%Control: year (1) truncated
%Control: production of eprint (0) enabled
%

\end{document}